\documentclass[a4paper,fleqn,usenatbib]{mnras}

\usepackage[T1]{fontenc}
\usepackage{ae,aecompl}

\usepackage{graphicx}	

\usepackage{amsmath}	
\usepackage{amssymb}	
\usepackage{enumerate,multirow}

\title[Correlations of the feedback energy and BCG radio luminosity in galaxy clusters] 
{Correlations of the feedback energy and BCG radio luminosity in galaxy clusters}
\author[Iqbal et~al.]
{Asif Iqbal$^{1}$\thanks{
asif@rri.res.in}, Ruta Kale$^2$\thanks{ruta@ncra.tifr.res.in}, Biman B. Nath$^{1}$\thanks{
biman@rri.res.in}, Subhabrata Majumdar$^{3}$\thanks{
subha@tifr.res.in} \\ 
  $^{1}$Raman Research Institute, Sadashiva Nagar, Bangalore, 560080, India\\
  $^{2}$National Centre for Radio Astrophysics, Pune, India\\
  $^{3}$Tata Institute of Fundamental Research, 1 Homi Bhabha Road, Mumbai, 400005, India\\
}
\date{Submitted to MNRAS Letters}
\pubyear{2018}

\begin{document}
\label{firstpage}
\pagerange{\pageref{firstpage}--\pageref{lastpage}}
\maketitle
\begin{abstract}
We study the excess entropy and the corresponding non-gravitational feedback energy ($E_{feedback}$) in the intra-cluster medium (ICM) by considering a sample of 38  galaxy clusters using  Chandra X-ray and  NRAO VLA Sky Survey (NVSS)/Giant Metrewave Radio Telescope (GMRT) radio observations. We find moderate correlation of the feedback energy and brightest cluster galaxy (BCG) radio luminosity ($L_R$) with the various cluster thermal properties. We show conclusively that the active galactic nucleus (AGN) is more effective in transferring feedback energy to the ICM in less massive clusters.  We find that within $0.3r_{500}$, the feedback energy correlates with cluster temperature as $E_{feedback}\propto T_{obs}^{0.98\pm0.37}$. Moreover, for radio detected BCG sample we find that BCG radio luminosity at 1.4 GHz scales with gas mass as $L_R\propto m_{g,obs}^{ 1.76\pm0.71}$ and with X-ray luminosity as $L_R\propto L_{X,obs}^{0.94\pm0.35}$. Finally, we discuss the implications of our results with regard to feedback in clusters.
\end{abstract}



\section{Introduction}
Galaxy clusters which grow through mergers in hierarchical structure formation play an important role in astrophysics and cosmology \citep{1}.  The total mass of the galaxy clusters comprises of main dark matter component  ($\approx$85\%), hot ICM ($\approx$10\%)  and remaining in the form of stellar matter, all of which are studied directly/indirectly with the help of X-ray, optical and gravitational lensing observations \citep{10,Bartelmann2001}. 

In galaxy clusters, a convenient way of describing the thermodynamical properties of the ICM is through the entropy which is usually defined as $K_g(r)=k_BTn_e(r)^{-2/3}$, where $k_B$ is the Boltzmann constant, $n_e$ is the electron number density and $T$ is the temperature of ICM. By knowing the entropy distribution and the total mass distribution, one can determine the density/temperature of the ICM using a hydrostatic equilibrium equation with a suitable boundary condition \citep{Nath2011}. 

Several observations have found higher gas entropy \citep{10,Eckert2013} than predicted by non-radiative hydrodynamical simulations \citep{Voit2005}, especially, near cluster centers. It has now become clear that various complex non-gravitational processes like feedback from AGN, radiative cooling and supernovae play  a vital role in modifying  the thermal structure of ICM \citep{Nath2002,Roychowdhury2005,Chaudhuri2012,Iqbal2017a}.  Investigation of non-thermal phenomena from radio observations and simulations has revealed radio mode AGN feedback based on bubble injection as a dominant role in adding feedback energy \citep{McNamara2007,Gaspari2014}.
 
The excess entropy and the corresponding feedback energy can be estimated by comparing the observed thermodynamic quantities  with that of the theoretical non-feedback (non-radiative) model \citep{Chaudhuri2012,Iqbal2017a}. In particular, \cite{Chaudhuri2012,Chaudhuri2013}  using XMM-Newton data found mean energy per particle to be $2.74\pm0.87$ keV up to $r_{500}$. Similarly, \cite{Iqbal2017a,Iqbal2017b} showed that feedback profiles become consistent with zero in the cluster outer regions ruling out pre-heating scenarios.

In this {\it letter}, we use a sample of 38 galaxy clusters having both Chandra X-ray data from the ACCEPT sample of \cite{Cavagnolo2009}\footnote{https://web.pa.msu.edu/astro/MC2/accept/clusters/.}  and NVSS/GMRT radio data from \cite{Kale2015a} to quantify the correlations of the energy feedback and BCG radio luminosity with related cluster bulk properties. In particular, we show that  AGN feedback is more efficient in less massive clusters. Unlike previous analysis of \cite{Chaudhuri2013}, who used average radio fluxes of all the sources near the cluster center, we use radio data from the optically identified BCGs to study the correlations.  We adopt a cosmology with $H_0=70$ km s$^{-1}$ Mpc$^{-1}$, $\Omega_M=0.3$ and $\Omega_\Lambda =0.7$.

\begin{table}
\caption{Basic properties of the clusters sample.}
\noindent\resizebox{\linewidth}{!}{
\centering
 \label{sample}
 \begin{tabular}{lccccc}
  \hline
Cluster &state&$z$  &$T_{obs}$ &$M_{500}$     & $L_R$   \\
 $-$ &$-$& $-$   & keV &$10^{14}M_{\odot}$  &$10^{38}$keV s$^{-1}$ Hz$^{-1}$   \\
  \hline 
ABELL 0068        &NCC   &    $0.25$ & $7.99$&$6.19$     &  $<26.6$    \\
ABELL 0141        &NCC   &    $0.23$ & $8.90$&$4.47$     &  $9.7\pm0.3$ \\
ABELL 0209        &NCC   &    $0.20$ & $8.28$&$8.17$     &  $<16.7$  \\
ABELL 0267        &NCC   &    $0.22$ & $6.79$&$4.94$     &  $<21.3$   \\
ABELL 0521        &NCC   &    $0.24$ & $6.74$&$6.90$     &  $2.7\pm0.4$   \\
ABELL 0611        &CC   &    $0.28$ & $6.69$&$5.85$     &  $7.8\pm7.0$  \\
ABELL 0697        &NCC  &    $0.28$ & $9.06$&$11.48$    &  $11.9\pm0.5$   \\
ABELL 0773        &NCC   &    $0.21$ & $8.53$&$7.08$     &  $<18.7$    \\
ABELL 0963        &CC   &    $0.20$ & $6.60$&$5.73$     &  $49.8\pm3.3$    \\
ABELL 1423        &CC   &    $0.21$ & $8.50$&$6.08$     &  $<17.9$    \\
ABELL 1576        &CC   &    $0.30$ & $8.65$&$5.98$     &  $351.4\pm7.8$    \\
ABELL 1758        &NCC   &    $0.27$ & $7.95$&$7.99$     &  $111.0\pm1.1$    \\
ABELL 1763        &NCC   &    $0.22$ & $6.90$&$8.29$     &  $7422.7\pm4.1$         \\
ABELL 1835        &CC   &    $0.25$ & $7.65$&$8.46$     &  $376.895\pm5.2 $        \\
ABELL 2111        &NCC   &    $0.22$ & $8.02$&$5.45$     &  $<21.1$     \\
ABELL 2163        &NCC   &    $0.20$ & $12.12$&$16.44$   &  $<16.1$     \\
ABELL 2219        &NCC   &    $0.22$ & $9.81$&$11.00$    &  $<20.9$     \\
ABELL 2261        &CC   &    $0.22$ & $7.58$&$7.38$     &  $66.3\pm4.0$    \\
ABELL 2390        &CC   &    $0.23$ & $9.16$&$9.48$     &  $2096.4\pm4.3$    \\
ABELL 2537        &CC   &    $0.29$ & $6.08$&$6.16$     &  $<37.8$    \\
ABELL 2631        &NCC   &    $0.27$ & $9.60$&$6.96$     &  $<32.6$    \\
ABELL 2667        &CC   &    $0.22$ & $6.31$&$6.81$     &  $187.2\pm4.1$                  \\
ABELL 2744        &NCC   &    $0.30$ & $9.61$&$9.55$     &  $<40.7$     \\
ABELL 2813        &NCC   &    $0.29$ & $8.39$&$9.16$     &  $<36.6$     \\
ABELL 3088        &CC   &    $0.25$ & $6.71$&$6.70$     &  $3.7\pm0.4$    \\
MACS J1115.8+0129 &CC   &    $0.34$ & $9.26$&$6.36$     &  $ 437.8\pm11.0$                  \\
MACS J1023.8-2715 &NCC  &     $0.30$&$8.43$& $8.83$      & $349.1\pm1.4$            \\
MACS J2211.7-0349 &CC   &    $0.39$ &$10.51$&$9.20$     &  $ <30.5         $         \\
MACS J2228+2036   &NCC   &    $0.41$ & $8.40$&$7.81$     &  $ <83.0         $        \\
MS   1455.0+2232  &CC   &    $0.25$ & $4.51$&$6.20$     &  $ <57.5\pm 5.5 $                \\
RX J0439.0+0715   &CC   &    $0.24$ & $6.50$&$5.74$     &  $7.9\pm0.3$    \\
RX J1504.1-0248   &CC   &    $0.21$ & $8.90$&$6.97$     &  $343.8\pm3.6  $                 \\
RX J1532.9+3021   &CC   &    $0.36$ & $5.44$&$9.50$     &  $372.6\pm10.6 $                  \\
RX J2129.6+0005   &CC   &    $0.23$ & $6.10$&$4.23$     &  $260.8\pm4.4  $                 \\
ZwCl 0857.9+2107  &CC   &    $0.23$ & $12.10$&$3.10$    &  $105.8\pm4.4  $                 \\
ZWCL 1953         &NCC   &    $0.37$ & $14.50$&$7.39$    &  $<63.9      $                \\
ZWCL 3146         &CC   &    $0.28$ & $12.8$&$5.30$     &  $52.7\pm7.1   $                \\
ZWICKY 2701       &CC   &    $0.21$ & $4.44$&$4.00$     &  $124.1\pm3.6  $                 \\

\hline
 \end{tabular}
 }
 Columns (1), (2),  (3), (4), (5) and (6) shows name, state, redshift, average temperature within the observed radius, $m_{500}$ and BCG radio luminosity at 1.4 GHz respectively. 
\end{table} 

\section{Cluster-BCG sample}
We started with the parent sample of BCGs identified in galaxy clusters in the Extended GMRT Radio Halo Survey (EGRHS) \citep{Kale2015a}. The EGRHS sample consists of clusters in the redshift range 0.2--0.4 that have X-ray luminosities, $\rm{L_{X[0.1-2.4keV]}} > 5 \times 10^{44}$ erg s$^{-1}$ and declinations $>-31^{\circ}$ \citep{Venturi2008, Kale2015b}. Only those clusters from EGRHS that were present in the ACCEPT sample were selected for this study. This led to a final sample of 38 clusters\footnote{RX J0439.0+0520 and RXC J1023.8-2715  for which the data is not up to $0.3r_{500}$ and Abell 520  for which there is no dominant galaxy that can be considered BCG \citep{Kale2015a,Kale2015b} were excluded.} with their corresponding BCGs as shown in Tab.~\ref{sample}: 23 with confirmed radio detected BCGs and 15 with upper limits to the radio powers (radio non-detection BCGs). 
The ``cool-core"  (CC) or ``non-cool-core" (NCC)  classification of the dynamical state of the cluster as used in \cite{Kale2015b} is given in column 2 of Tab. \ref{sample}. This classification is based on the X-ray morphological parameters, namely,  power ratio ($P_3/P_0$), centroid shift ($w_{500}$) and concentration ($c_{100}$), 
that are described in \citet{Cassano2010}. They classify a cluster as NCC if $P_3/P_0 > 1.2 \times10^{-7}$, $w_{500} >0.012$ and $c_{100}<0.20$. It is important to note that radio detection sample is dominated by the cool-core (CC) clusters (17 out of 20) while as non-detection sample is dominated by non cool-core (NCC) clusters (12 out of 18). The 1.4 GHz radio powers of the BCGs from  \citet{Kale2015a} included the K-correction. A spectral index\footnote{The spectral index, $\alpha$ for a synchrotron spectrum is defined as $S_\nu \propto \nu^{-\alpha}$, where $S_\nu$ is the flux density at frequency $\nu$.} of 0.8 for the radio continuum spectra of the BCGs was assumed. For the BCGs that were not detected in radio bands, the upper limits at 1.4 GHz correspond to five times the rms noise ($5\times0.45$ mJy beam$^{-1}$) in the NVSS \citep{Condon1998}. 
\begin{figure*}                                                                                                                                                                                                                                                                                                                                                                                                                                                                                                                                                                                                                  
\includegraphics[width=0.45\linewidth,height=4.5cm]{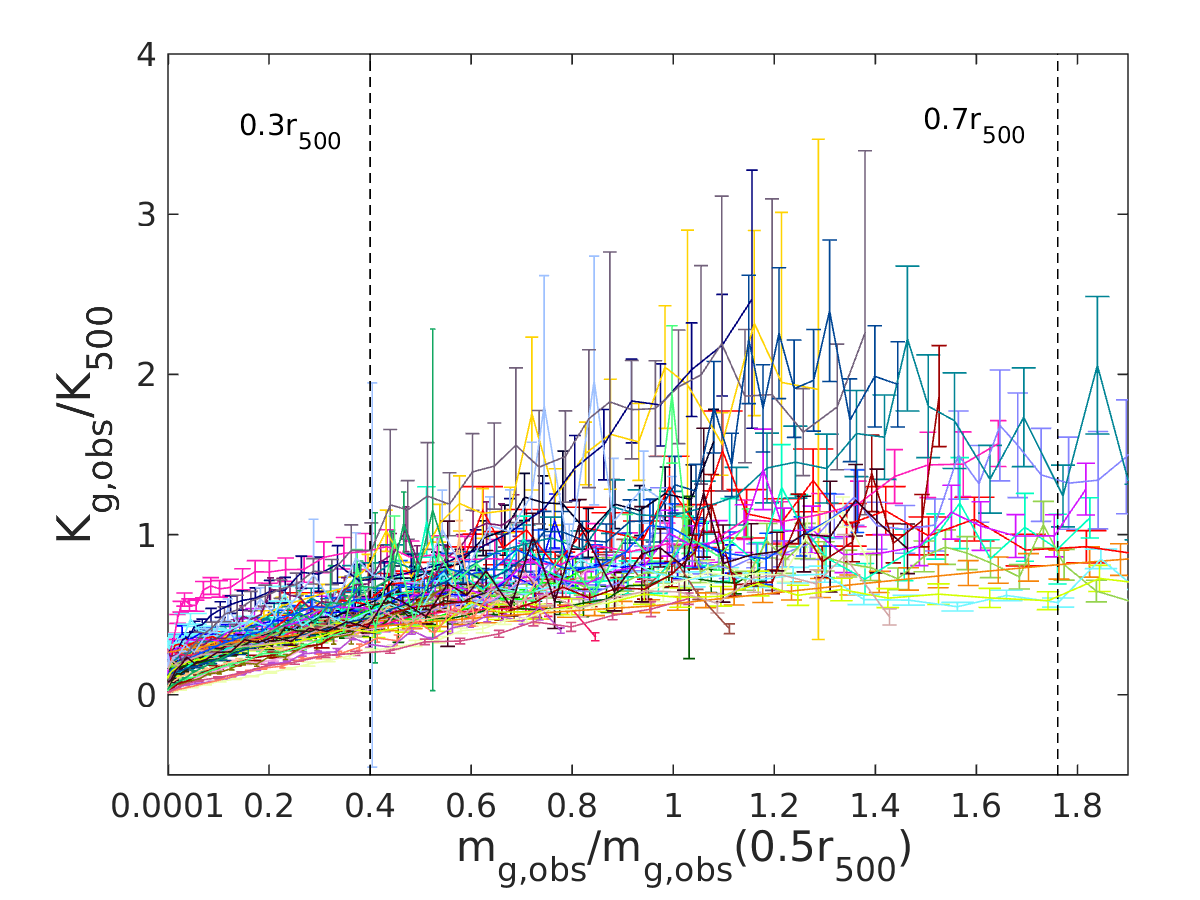}
\includegraphics[width=0.45\linewidth,height=4.5cm]{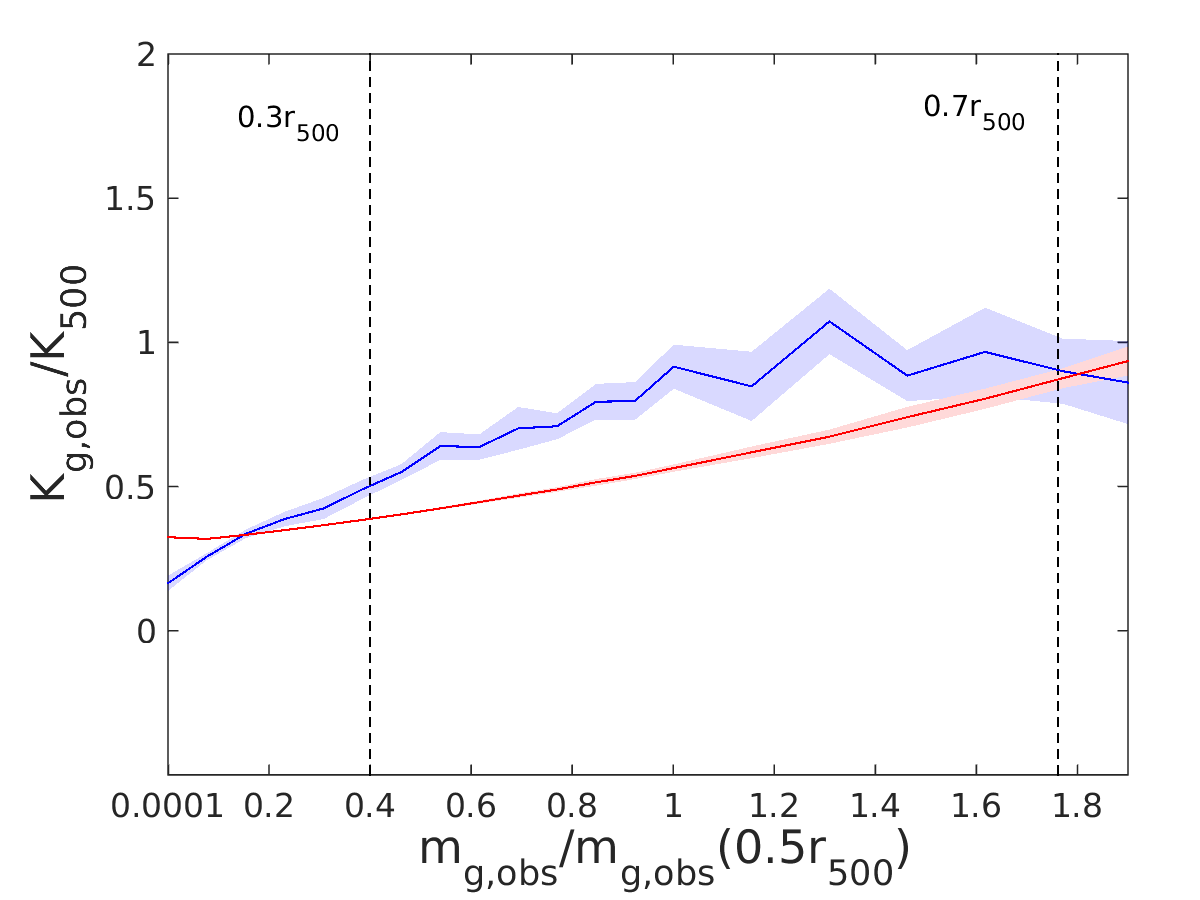}
\caption{Left-hand panel: Observed entropy  as a function $m_g/m_g(0.5r_{500})$  for all the clusters. Right-hand panel: 
Comparison of the median observed and theoretical entropy profiles \citep{Voit2005} as a function $m_g/m_g(0.5r_{500})$. The error bars are at $1-\sigma$ confidence.}
\label{figure1}
\end{figure*}
\section{Non-radiative Model of ICM}
The ICM is taken to sit in the gravitational potential of the dark matter halo having a Navarro-Frenk-White (NFW) density profile \citep{Navarro1996} characterized by $\rho_{\textrm{dm}}=\frac{\rho_s}{x(1+x)^2}$,
where $x=r/r_s,$ $r_s$ is the scale radius and $\rho_s$ is the normalization factor in units of density. The concentration parameter is given by $c_{\Delta}=r_{\Delta}/r_s$, where 
$\Delta$ is defined such that $r_{\Delta}$ is the radius out to which mean matter density is $\Delta \rho_c(z)$, $\rho_c(z)$ being critical density of the universe at redshift $z$. 
We use $m_{500}$ from \cite{Planck2013}\footnote{http://szcluster-db.ias.u-psud.fr.} with the exception of clusters RX J1532.9+3021 (MACS J1532.8+3021), ZwCl 0857.9+2107 (ZWICKY 2089) and ZWICKY 2701 whose values were taken from \cite{Mantz2010}. Further, we fix the NFW concentration parameter to be $c_{500}=3.2$ \citep{Pointecouteau2005,10}.
The viral radius, $r_{vir}$, is calculated with spherical collapse model,
 $r_{vir}=\left[\frac{m_{vir}}{4\pi/3\Delta_c(z)\rho_c(z)}\right]^{1/3}$, where $\Delta_c(z)=18\pi^2+82(\Omega_M(z)-1)-39(\Omega_M(z)-1)^2$ \citep{Bryan1998}.
\begin{figure*}
\includegraphics[width=0.45\linewidth,height=4.5cm]{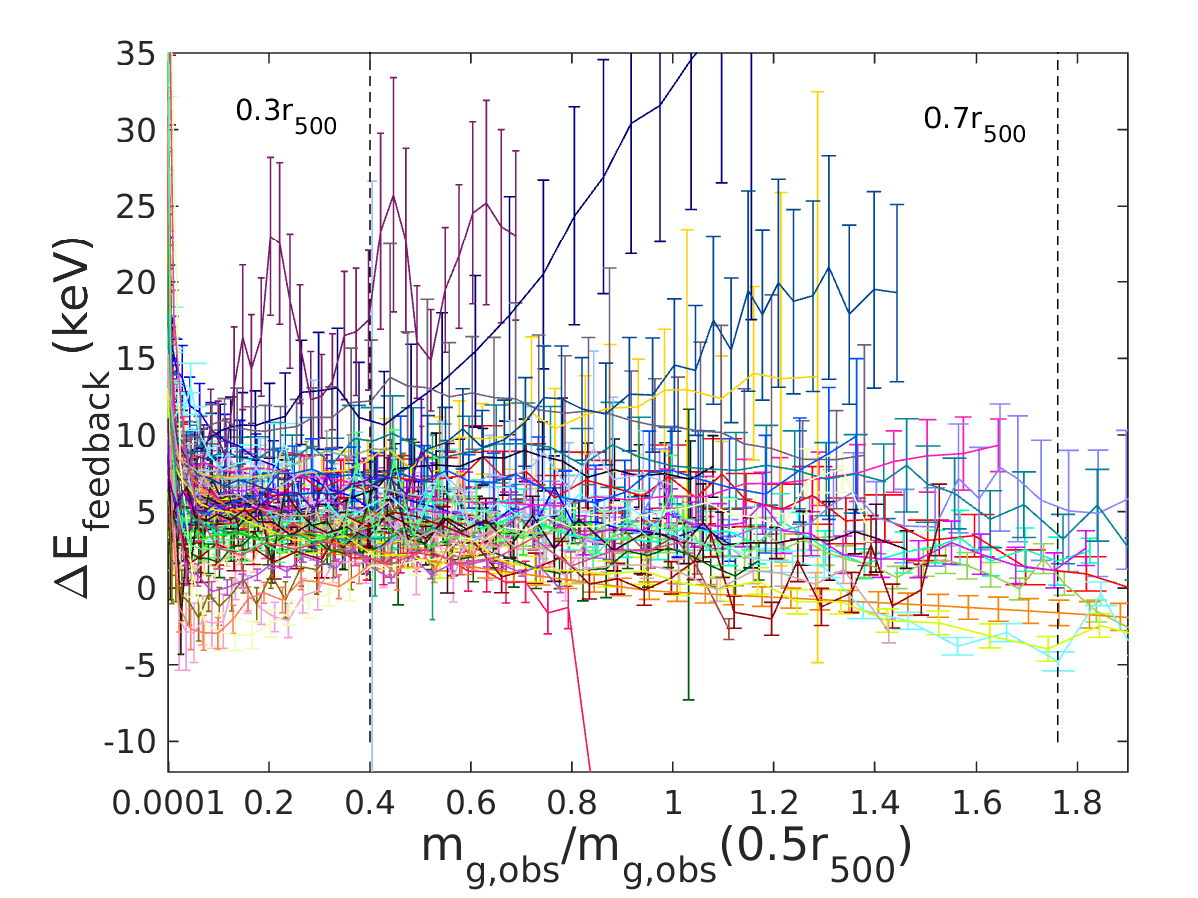}
\includegraphics[width=0.45\linewidth,height=4.5cm]{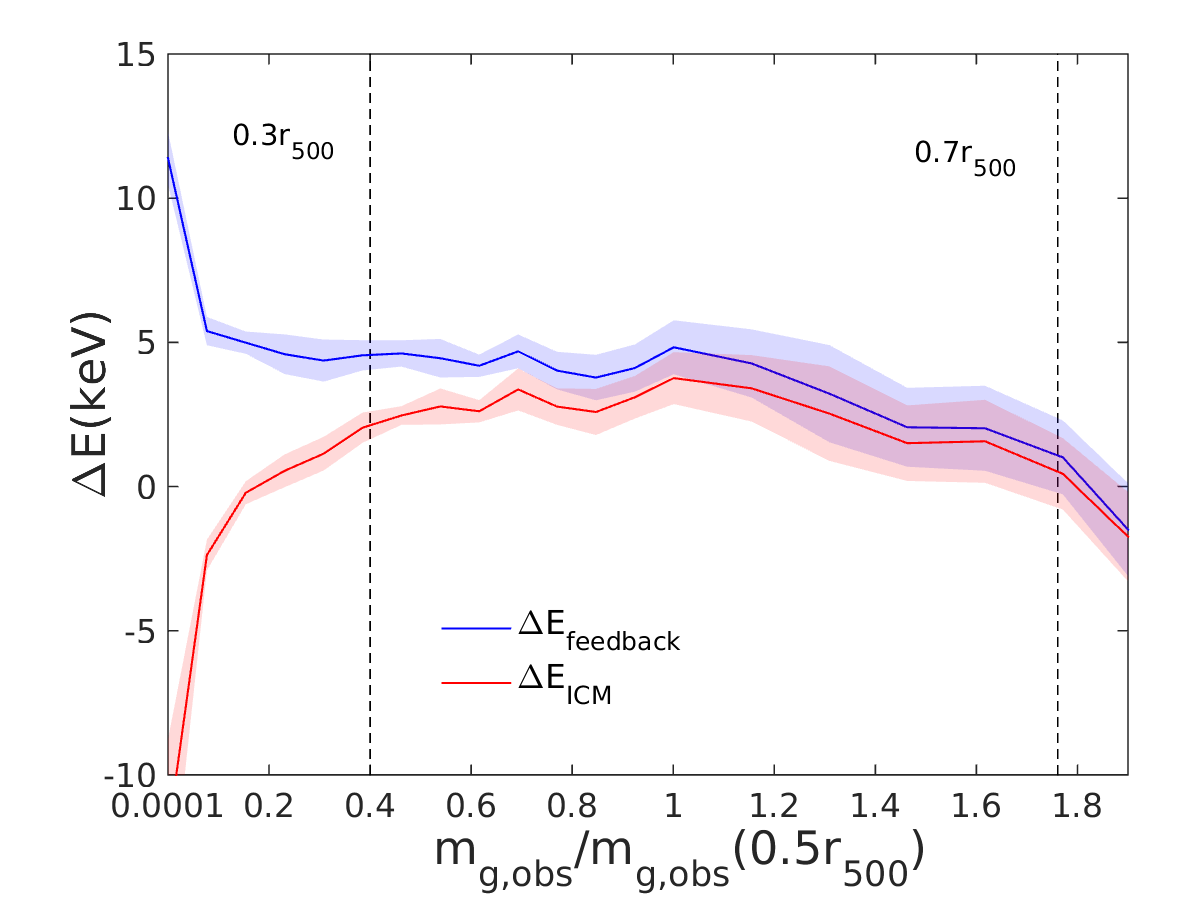}
\caption{Left-hand panel: Non-gravitational energy per particle as a function $m_g/m_g(0.5r_{500})$  for all the clusters. Right-hand panel: Comparison of the median of $\Delta E_{feedback}$ with that of $\Delta E_{ICM}$. The error bars are at $1-\sigma$ confidence.}
\label{figure2}
\end{figure*}

\cite{Voit2005} using non-radiative AMR and SPH simulations observed that entropy profiles scales as $K_{g,th}\propto r^{1.1-1.2}$ in the range $(0.2-1)r_{200}$ and flatten in the cluster cores.  They found differences in the entropy profiles in cluster cores between AMR and SPH. However, it is now clear that the two results become consistent with one another after accounting for shocks and mixing motions in the SPH case \citep{Mitchell2009,Vazza2011}. We therefore  use the AMR median entropy profile obtained by \cite{Voit2005} and fit it with an appropriate fourth order polynomial in the whole radial range \citep{Chaudhuri2013},
\begin{eqnarray}
 \frac{K_{g,th}(r)}{K_{200}}= \sum_{i=0}^{4}  a_i\left(\frac{r}{r_{200}}\right)^i,
\end{eqnarray}
where $K_{200}=144 \left(\frac{m_{200}}{10^{14}M_\odot}\right)^{2/3} \left(\frac{1}{f_b}\right)^{2/3} h(z)^{-2/3} \mbox{ keV cm}^2$, $f_b$ being the universal baryonic fraction, $h(z)=H(z)/H_0$ and $a_0$=$0.193$, $a_1$=$-0.375$, $a_2$=$3.850$, $a_3$=$-3.080$, $a_4$=$0.868$. 

The  gas density ($n_{g,th}$) and temperature ($T_{th}$) profiles for theoretical model are obtained by numerically solving hydrostatic equation  with an appropriate boundary condition given by  $f_g=0.156$ \citep{Planck2015} at the virial radius \citep{Chaudhuri2012,Chaudhuri2013},
\begin{equation}
\frac{dP_{g,th}(r)}{dr} =-\left[ \frac{P_{g,th}(r)}{K_{g,th}(r)} \right ]^{3/5} m_p \mu_e^{2/5} \mu^{3/5} \frac{ G M( <r ) }{ r^2} ,
\label{he2}
\end{equation}
where $P_{g,th} = n_{g,th} k_B T_{th}$ is the theoretical pressure of ICM, and $M( <r )$ is the total mass of cluster within radius $r$. The left-hand panel in the Fig.~\ref{figure1} shows the individual observed cluster entropy profiles as a function of gas mass $m_g$ while the right-hand panel shows the observed median  profile and \cite{Voit2005} theoretical median entropy profile for the whole sample.  Since entropy is a Lagrangian quantity, we compare the profiles at same gas mass instead of same radii in order to take into account redistribution of gas due to feedback \citep{Nath2011,Chaudhuri2012,Iqbal2017a}.
At a given mass shell the median profiles where obtained using 1000 bootstrap iterations by means of re-sampling of data points with repetitions. The errors bars are then given by root mean square deviation of the distribution. The vertical lines in the Fig.~\ref{figure1} approximately define the  core region ($r<0.3r_{500}$) and outside core region ($0.3r_{500}\leq r \leq 0.7r_{500}$). As can be seen there is entropy excess up to $0.7r_{500}$ except at the very centers where the high degree of radiative loss has resulted in the observed entropy being less than theoretical one.
\section{Estimates of feedback profiles}
The amount of thermal energy deposition  is found to be proportional to $T_{obs}\Delta K/K_{obs}$, where $\Delta K = K_{obs}-K_{th}$. Considering isobaric process, the additional non-gravitational thermal energy per particle in ICM is,
\begin{eqnarray}
\Delta Q_{ICM} &=&{k_BT_{obs}  \over (1-{1 \over \gamma})} 
{  \beta ^{2/3} (\beta -1) \over (\beta^{5/3}-1)}
{\Delta K \over K_{g,obs}},
\label{eq:delq}
\end{eqnarray}
where $\beta=T_{obs}/T_{th}$.  The excess energy per particle
is then obtained by adding the change in potential energy in Eq.~\ref{eq:delq},
\begin{equation}
\Delta E_{ICM}=\Delta Q_{ICM}+G\mu m_p\left(\frac {M_{tot}(r_{th})}{r_{th}}-\frac{M_{tot}(r_{obs})}{r_{obs}}\right), 
\end{equation}
where $r_{th}$ and $r_{obs}$ are theoretical and observed radii respectively enclosing the same gas mass.

Finally, the total feedback energy/particle can be found after adding the energy lost due to radiative cooling,
\begin{equation}
\Delta E_{feedback}= \Delta E_{ICM}+ \Delta L_{X}\,t_{age},
\label{energy}
\end{equation}
where $\Delta L_{X}$ is the bolometric luminosity emitted by the ICM in a given shell which is estimated by averaging theoretical and observed cooling function, $\Lambda_{N}$  of \cite{Tozzi2001}. $t_{age}$ is the age of the shell which is calculated using the expression of mass acceration rate given by \cite{Voit2003} for clusters of present day mass $\approx10^{15}M_{\odot}$ and taking the age of Universe to be 13.47 Gyrs.  The total excess energy deposited within the radius $r$ is given by  $E_{feedback} \,= \,\frac{1}{\mu_g m_p} \,\int_0^{r} \Delta E_{feedback} \, dm_g \,$,
where $\mu_g = 0.6$ is the mean molecular weight of gas and $m_p$ is mass of proton. The average energy/particle ($\epsilon_{feedback}$) is found by dividing $E_{feedback}$  with the total number of gas particles $N(r)$. 

In Fig.~\ref{figure2}, the left-hand panel shows the non-gravitational energy profiles as a function of gas mass $m_g$ for individual clusters and the right-hand panel shows the median non-gravitational feedback energy with and without adding energy lost due to cooling.  In par with earlier findings of \cite{Chaudhuri2012,Iqbal2017a}, our results also find significant entropy and hence evidence of feedback energy in the cluster inner regions. Moreover, as can be seen the radiative loss is only important up to $0.3r_{500}$.  We find that the average feedback energy per particle $\epsilon_{feedback}$ to be $4.32\pm0.52$ keV in the  region $(0.01-0.3)r_{500}$ and $4.54\pm0.55$ keV in the region $(0.01-0.5)r_{500}$.

\begin{figure}
\includegraphics[width=.90\linewidth,height=4.5cm]{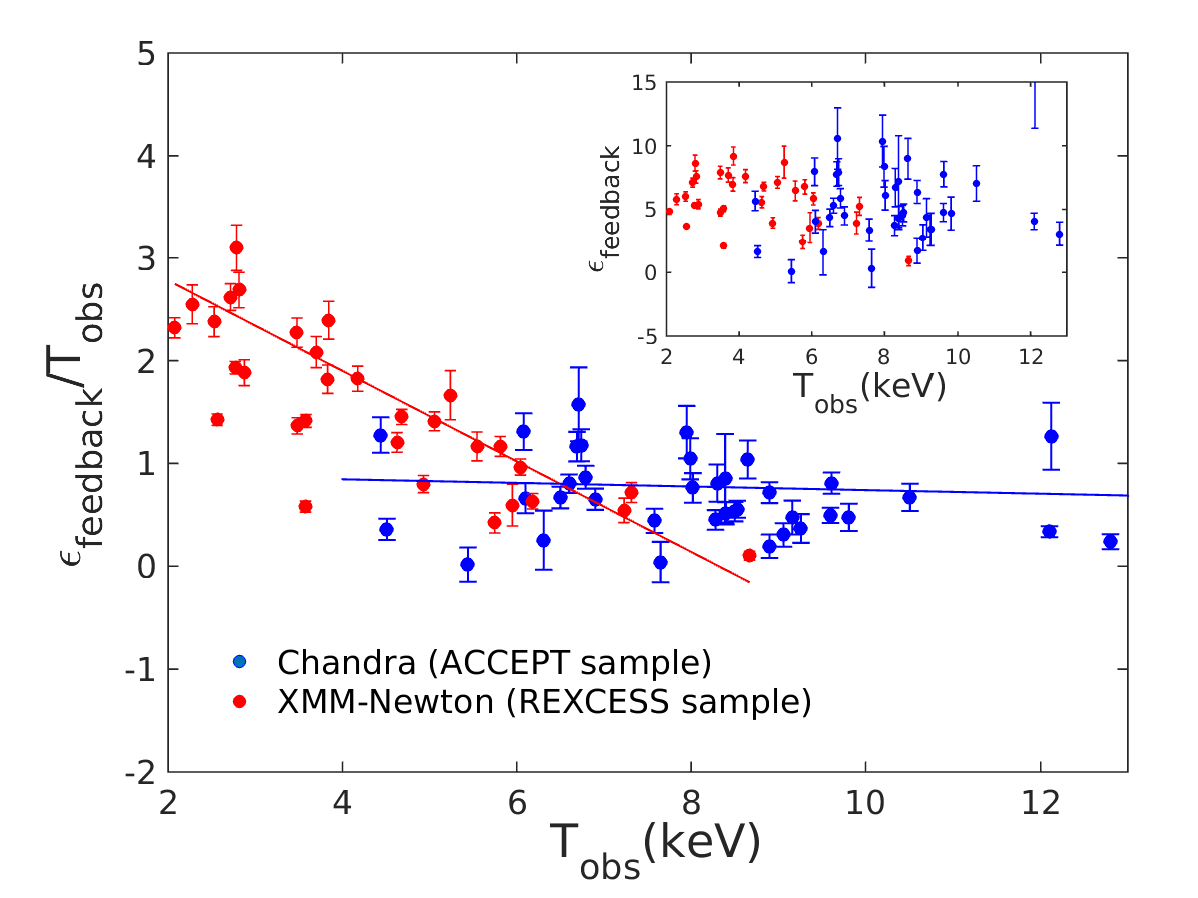}
\caption{Comparison of $\epsilon_{feedback}/T_{obs}$ between REXCESS sample \citep{10} and our Chandra sample of high temperature clusters ($>6$keV). Solid blue and red lines show the best-fit for Chandra and REXCESS samples respectively. The inset shows the feedback energy per particle.}
\label{figure3}
\end{figure}

Fig.~\ref{figure3} shows the ratio of the non-gravitational feedback energy over energy from gravitational collapse, $\epsilon_{feedback}/T_{obs}$ as a function of $T_{obs}$ in the region $(0.01-0.3)r_{500}$ for our sample sample and for REXCESS sample\footnote{We recalculated the values in \cite{Chaudhuri2013} for REXCESS sample including our Eq.~\ref{energy}.}. It can be clearly seen that there is a higher degree of feedback for the REXCESS sample which are mainly low temperature clusters (<6keV) compared to our sample which are high temperature  clusters (>6keV). This shows that for low temperature (mass) clusters, the AGN feedback is more effective in transferring energy into ICM.  A simple linear fitting of $\epsilon_{feedback}/T_{obs}=b_1T_{obs}+b_0$ yields $b_1=-0.01\pm0.03$, $b_0=0.915\pm0.30$ for our sample and $b_1=-0.44\pm0.05$, $b_0=3.66\pm0.23$ for REXCESS sample. Our result corroborates  previous works on the non-gravitational feedback \citep{Fabian2012,McNamara2007}.

\begin{table}
\caption{Best-fit scaling relations and Spearman's rank correlation coefficient ($r$).}
\noindent\resizebox{\linewidth}{!}{
 \centering
\begin{tabular}{c c c c c c c}
 \hline
Method  & $A$ &  $B$ & $r$ & $A$ &  $B$ & $r$\\
\hline
&\multicolumn{3}{c}{Full sample}& \multicolumn{3}{c}{Radio detection only}\\
\hline 
 \multicolumn{7}{c}{$\log(E_{feedback}/10^{70}keV) = A + B\log(T/8keV)$}\\
\hline 
EM        &$1.06\pm0.05$   &$1.56\pm0.47$ & $0.52$    &  $-$  &$-$&  $-$  \\
Bayesian  &$1.12\pm0.05$   &$0.98\pm0.37$ &  $-$         &  $-$  &$-$& $-$  \\

 \hline

  \multicolumn{7}{c}{$\log(L_R/10^{38}keV s^{-1}Hz^{-1}) = A + B\log(m_{g,obs}/10^{13}M_{\odot})$}\\
\hline
EM       & $1.31\pm0.19$ &$1.17\pm0.71$&  $0.31$      &  $1.76\pm0.16$   &$1.78\pm0.61$&  $ 0.57$  \\ 
Bayesian & $1.60\pm0.13$ &$1.22\pm0.53$ &   $-$       &  $1.76\pm0.19$  & $1.76\pm0.71$& $-$  \\
\hline
 \multicolumn{7}{c}{$\log(L_R/10^{38}keV s^{-1}Hz^{-1}) = A + B\log(L_{X,obs}/10^{53}M_{\odot})$}\\
 \hline
EM  &         $0.81\pm0.31$ &$0.91\pm0.35$ &  $0.42$   &  $1.25\pm0.27$   &$0.95\pm0.30$&  $0.60$  \\ 
Bayesian &    $1.17\pm0.23$ &$0.81\pm0.27$ &   $-$        &  $1.27\pm0.32$  & $0.94\pm0.35$&  $-$ \\
\hline
\multicolumn{7}{c}{$\log(L_R/10^{38}keV s^{-1}Hz^{-1}) = A + B\log(M_{vir}/10^{14}M_{\odot})$}\\
 \hline
EM  &       $ 2.11\pm1.38$   &$-0.64\pm1.32$&  $ 0.02$    &$0.50\pm1.36$   &$1.43\pm1.32$&  $0.38$  \\ 
Bayesian  &  $1.53\pm1.00$  & $0.23\pm0.94$&   $-$           &$0.53\pm1.57$   &$1.41\pm1.53$&  $-$ \\ 
\hline
\end{tabular}
}
\label{tab:fit}
Note: $E_{feedback}$, $L_{X,obs}$ and $m_{g,obs}$ are estimated within $0.3r_{500}$.
\end{table}

\section{Correlations of the feedback energy and BCG radio luminosity}
Since the effect of AGN feedback is dominant only in the cluster inner regions \citep{Gaspari2014,Iqbal2017a}, we correlate clusters quantities measured within $r=0.3r_{500}$ (except for temperature) in order to gain meaningful picture AGN-ICM interaction. To estimate correlations we fit the power-law relations using linear regression in log-log space. The regression is first performed using parametric EM (Expectation Maximization) algorithm that is implemented in the ASURV package \citep{Isobe1986}. Since ASURV does not take errors into account, we  also consider Bayesian regression algorithm implemented in Linmix package\footnote{Python version - https://github.com/jmeyers314/linmix.} \citep{Kelly2007} which takes heteroscedastic and intrinsic scatter into account. However, both the algorithms incorporate upper limits.  We quote results from Linmix although both packages give similar results. To study the correlations of the BCG radio luminosity, we consider full sample as well as sub-sample of radio detected BCG clusters. Since detection sample is dominated by CC clusters  and non-detection sample is dominated by the NCC clusters it makes sense in separating the sample in these two groups. Tab. \ref{tab:fit} gives the best-fit results and correlation coefficient between various cluster parameters. In general, we find that for the radio detected sample, the best-fit lines have steeper slopes with larger values of correlation coefficient  compared to those from the full sample.

\begin{figure}

 \includegraphics[width=0.90\linewidth,height=4.5cm]{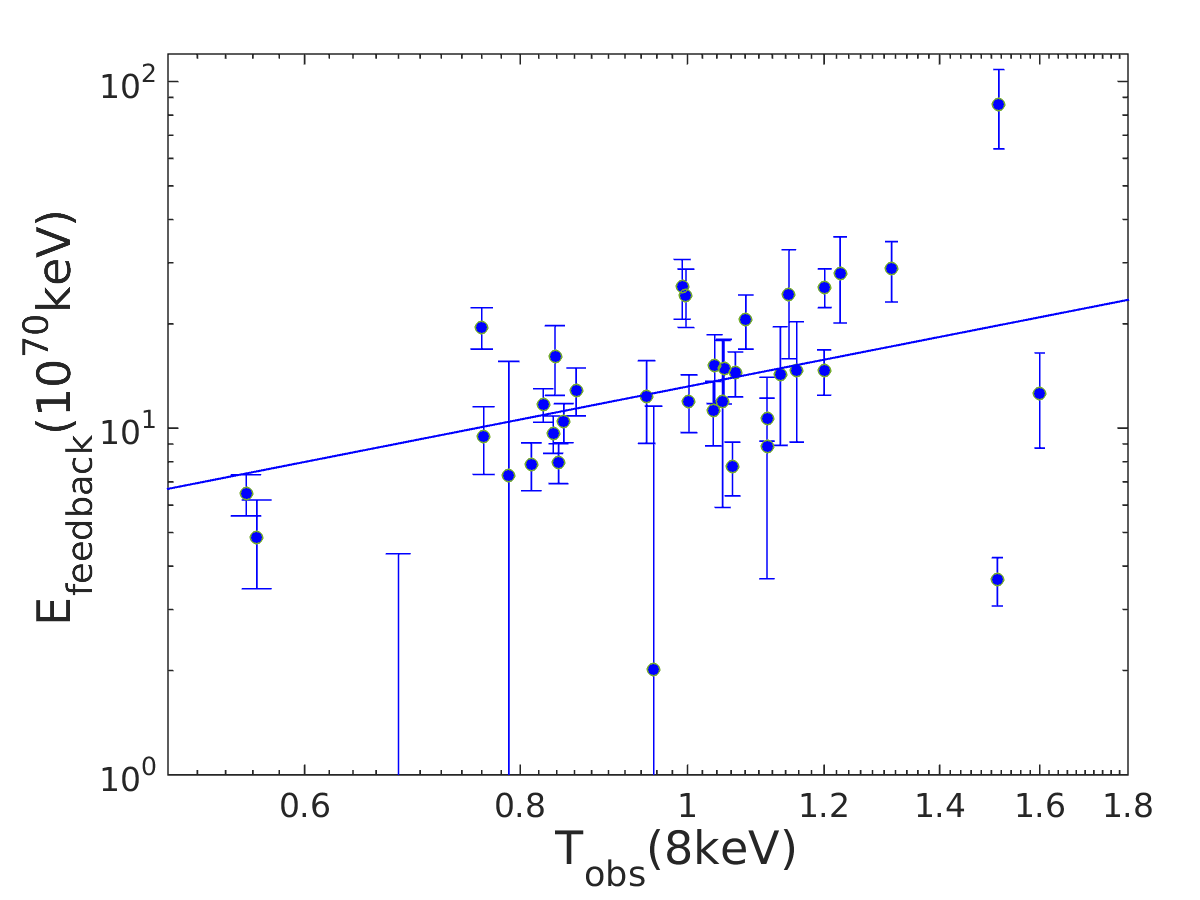}
\caption{Correlation between  $ E_{feedback}$ and $T_{obs}$. Solid blue line represent Bayesian best-fit.}
\label{figure4}
\end{figure}

\begin{figure*}
\begin{minipage}{5.8cm}
\includegraphics[width=5.8cm]{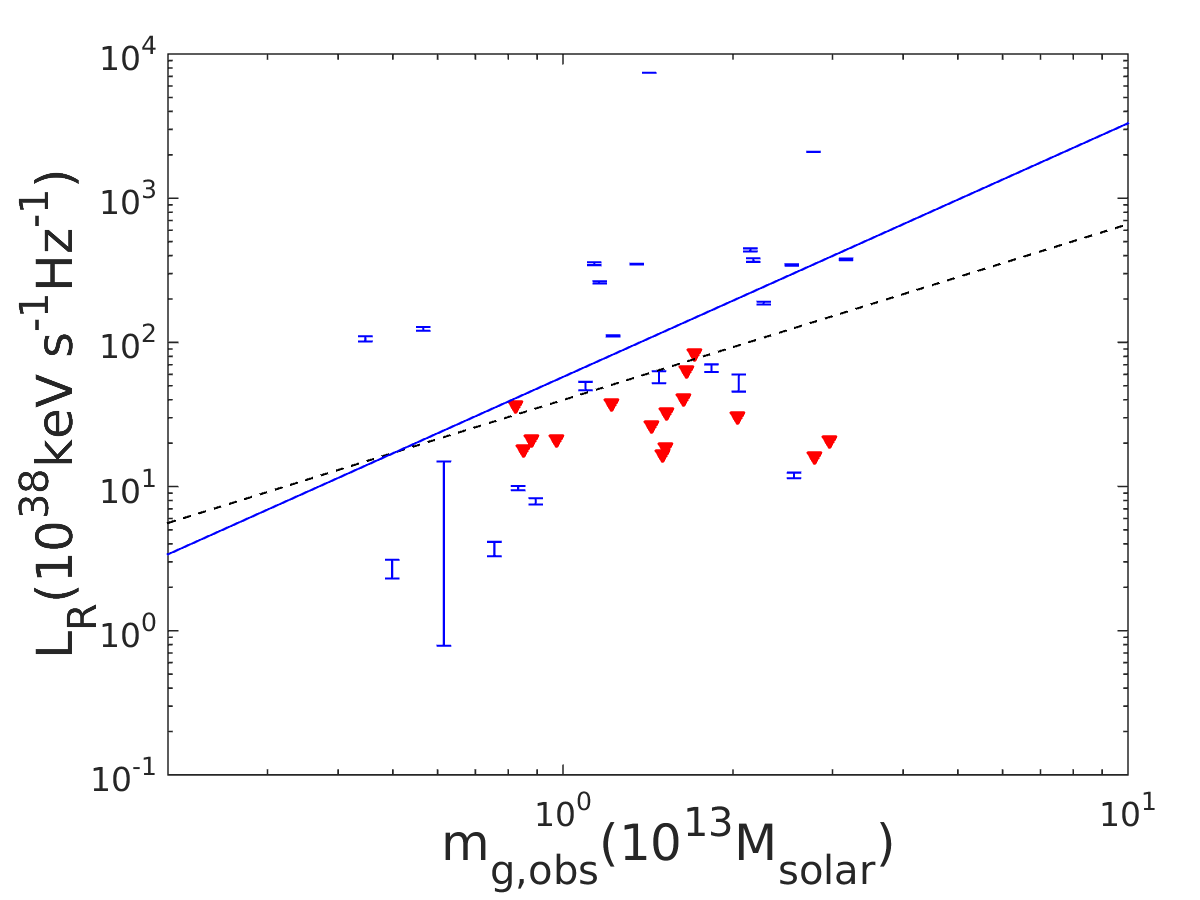}
\end{minipage} 
\begin{minipage}{5.8cm}
 \includegraphics[width = 5.8cm]{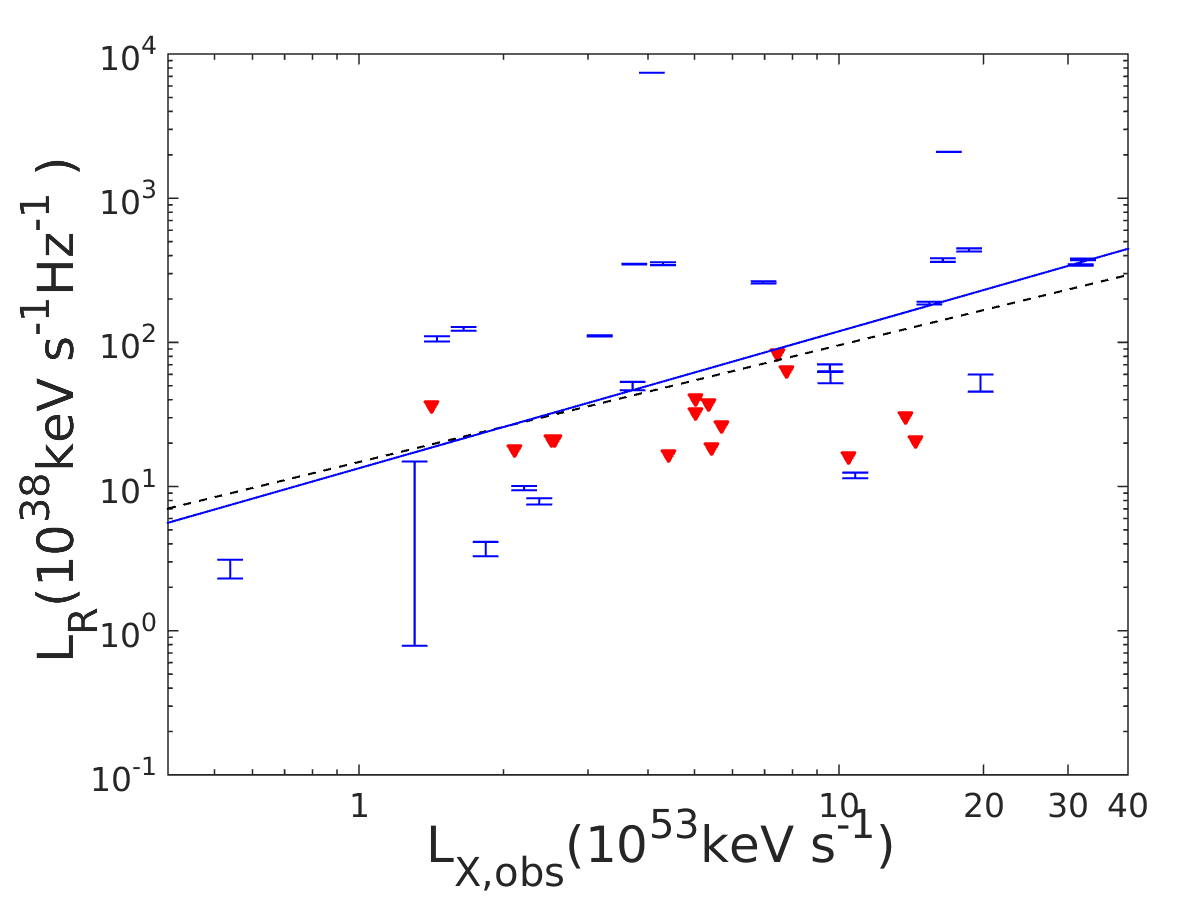}
\end{minipage} 
\begin{minipage}{5.8cm}
\includegraphics[width=5.8cm]{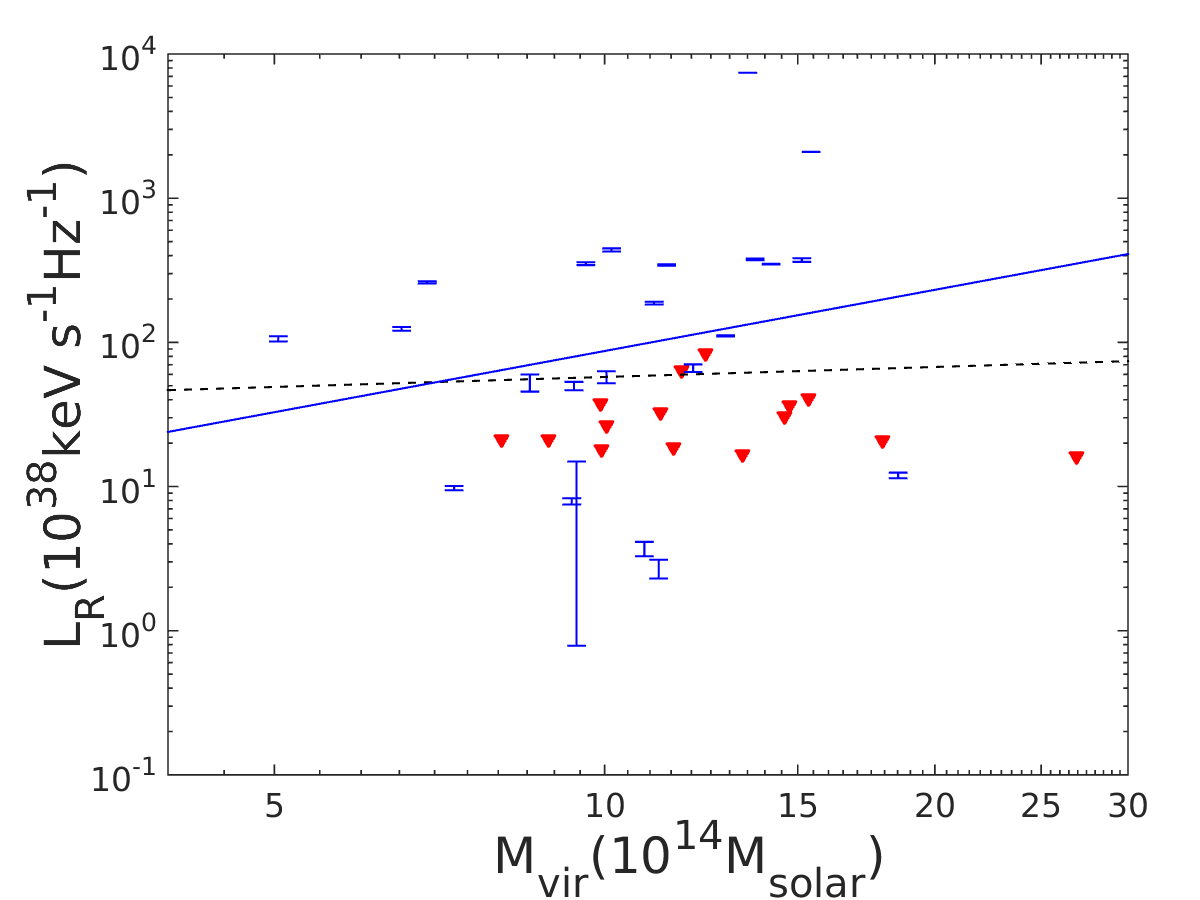}
\end{minipage}
\caption{Left-hand panel: Correlation between $L_R$ and $m_{g,obs}$. Middle panel: Correlation between $L_R$ and $L_{X,obs}$. Right-hand panel: Correlation between $L_R$ and $M_{vir}$. Solid blue and dashed black lines represent Bayesian best-fit for the full-sample and sub-sample (BCG detected) respectively.  Blue and red markers represent radio detection and radio non-detection samples respectively. }
\label{figure5}
\end{figure*}

Fig.~\ref{figure4}, shows the correlation between $E_{feedback}$ and $T_{obs}$. We find  $E_{feedback}\propto T_{obs}^{0.98\pm0.37}$ with a correlation coefficient of $0.52$ for the full-sample. This suggests that for massive clusters (high temperature),  although, as discussed in the previous section the fraction increase in energy per particle is small, the total feedback energy is large in massive clusters. Further, higher total mass also implies higher gas mass (or $N$) which makes feedback energy per particle ($\epsilon_{feedback}$) more or less constant for all temperature range (see inset in Fig~\ref{figure3}). From the self-similar consideration ($N\propto m_{g,obs} \propto M_{tot}\propto T_{obs}^{3/2}$), one obtains $\epsilon_{feedback}=E_{feedback}/N \propto T_{obs}^{0.5\pm0.37}$ which roughly agrees at $1-\sigma$ level with ours results.

The left-hand panel of Fig.~\ref{figure5} shows the correlation between $L_R$ and $m_{g,obs}$. We find the scalings $L_R\propto m_{g,obs}^{1.22\pm0.53}$ with the correlation coefficient of $0.31$ for full-sample and $L_R\propto m_{g,obs}^{1.76\pm0.71}$ with the correlation coefficient of $0.57$ for sub-sample. The middle-hand panel of Fig.~\ref{figure5} shows the correlation  $L_R$ with $L_{X,obs}$. We find $L_R\propto L_{X,obs}^{0.81\pm0.27}$ with the correlation coefficient of $0.42$ for the full-sample and $L_R\propto L_{X,obs}^{0.94\pm0.35}$ with the correlation coefficient of $0.60$ for sub-sample. This confirms the fact that  radio luminosity is proportional to the mass accretion rate which in turn depends on the gas mass and hence X-ray luminosity. 
 
Finally, the right-hand panel of the Fig.~\ref{figure5},  shows the correlation between the $L_R$ and $M_{vir}$ for which we obtain the poor estimates of the fitted parameter and weak correlation coefficient.  We find $L_R \propto M_{vir}^{0.23\pm0.94}$  with the correlation coefficient of $0.02$ for the full-sample and $L_R \propto M_{vir}^{1.41\pm1.53}$  with the correlation coefficient of $0.38$ for the sub-sample.

\section{Discussion and Conclusions}
Our finding of the above scaling between BCG radio luminosity $L_R$ and the cluster virial mass has important implications. It is also consistent with previously discovered scalings, as we will discuss below.

\cite{Franceschini1998} found that black hole mass ($M_{BH}$) in AGNs  scales with  radio luminosity as $L_R\propto M_{BH}^{2.5}$. 
There is also a relation between the total mass of a massive elliptical galaxy ($M_{BCG}$), $M_{BH}\propto M_{BCG}^{1.4}$ \citep{Reines2015}. 
Moreover, SDSS studies such as \cite{Behroozi2010} show that for such massive galaxies, the stellar mass $M_{\ast, BCG}$ scales as $M_{BCG}^{0.3}$. Combining these three scalings together, we find that, $L_R \propto M_{\ast, BCG}^{2.5 \times 1.4/0.3}= M_{\ast,BCG} ^{11.7}$. In addition,  $M_{*,BCG}$ scales with $M_{vir}$ of the parent cluster with a slope $0.12\pm0.03$ \citep{Whiley2008}. Using this, we obtain $L_R \propto M_{vir}^{1.4}$ which is consistent with our results for the sub-sample of radio detected BCG clusters\footnote{ If one instead uses slope of $0.42\pm0.07$ \citep{Chiu2016} (their Figure 2) for $M_{*,BCG}-M_{vir}$ relation then this gives $L_R \propto M_{vir}^{4.9}$.}.
Further combining $L_R\propto M_{BH}^{2.5}$ with our results of $L_R\propto M_{vir}^{1.29\pm1.59}$, one finds $M_{BH}\propto M_{vir}^\beta$, with $\beta=0.56^{+0.61} _{-0.60}$ for the sub-sample which is consistent with \cite{Roychowdhury2005} who found $M_{BH}\propto M_{vir}^{1}$ from excess entropy consideration.

Our finding that the feedback energy for a given radio luminosity decreases with increasing cluster mass (or temperature) also deserves attention. If the energy deposited by the radio source is through the dynamics of the cocoon, then one expects a constant fraction of the total energy of the radio source to be given as feedback energy, e.g., as derived by \cite{Bicknell1998} (their Eq. 2.13). Clearly, this is not tenable in light of our finding. However, it has been previously discussed in the literature that the efficiency of energy deposition may be larger for low mass clusters. \cite{Fabian2012} have suggested that weak shocks (expected in hot ICM of massive clusters) are poor at dissipating energy, and \cite{McNamara2007} suggested that a lower binding energy per particle in groups may lead to a greater efficiency of non-gravitational heating in low mass clusters. The high probability of radio detections in CC clusters suggest that it depends on the dynamical state of host cluster. Nevertheless, mergers may transform CC clusters into NCC clusters with enhanced ICM entropies. Alternatively, the lack of radio emission in the NCC clusters to account for the excess entropy suggest that clusters were pre-heated before cluster formation \citep{Dwarakanath2006}.

Since $L_R$ is directly linked with the thermal properties of the ICM, this motivates us to look for a relation between $E_{feedback}$  and $L_R$. For the current sample, we did not find any significant correlation between $E_{feedback}$  and $L_R$, probably because $E_{feedback}$ is the integrated quantity and $L_R$ is the current property. However, some possibility of separating the sample into the clusters where the radio emission is very recent (and hence not much  energy has been injected into ICM) and clusters where radio feedback has happened in the distinct past one might be able to find  interesting clues about $E_{feedback}$-$L_R$ relation and would be interesting extension of the work.

In summary, our study suggests that the non-thermal emission from the BCGs is directly linked  with the feedback energy and the thermodynamic properties of the ICM. We find moderate correlation of the feedback energy and BCG radio luminosity with the cluster properties. Our results suggest clusters which are radio detected and those without correlate differently with the ICM properties. Studies such as ours can be  powerful tool to  study the connection between BCGs (which are mostly found in CC clusters) and ICM thermodynamics and understanding dynamical/evolutionary differences of CC clusters from their NCC counterparts. Lastly, with the upcoming and future radio data such as from SKA, it will be possible to obtain the much tighter constraints on the scaling relations to better understand the effects of feedback on the cluster properties.  
\section*{Acknowledgements}
AI would like to thank Tata Institute of Fundamental Research (TIFR), Mumbai and National Centre for Radio Astrophysics (NCRA), Pune for hospitality. The authors would like to thank  anonymous reviewer for the critical review that have lead to significant improvement of the manuscript.

    \bibliographystyle{mn2e}

\label{lastpage}
\end{document}